\newlength{\abstwidth}
\documentclass[11pt,a4paper]{article}
\pdfoutput=1
\usepackage{graphicx}
\usepackage{hyperref}
\usepackage{amsmath}
\usepackage{amssymb}
\begin{document}
\sloppy
 
\pagestyle{empty}
 
\begin{flushright}
LU TP 15-57\\
MCnet-15-37\\
December 2015
\end{flushright}

\vspace{\fill}

\begin{center}
{\Huge\bf Hard Diffraction in \textsc{Pythia}~8\footnote{Presented at XLV International Symposium on 
Multiparticle Dynamics}}\\[4mm]
{\Large Christine O. Rasmussen} \\[3mm]
{\it Theoretical Particle Physics,}\\[1mm]
{\it Department of Astronomy and Theoretical Physics,}\\[1mm]
{\it Lund University,}\\[1mm]
{\it S\"olvegatan 14A,}\\[1mm]
{\it SE-223 62 Lund, Sweden}
\end{center}

\vspace{\fill}

\begin{center}
\begin{minipage}{\abstwidth}
{\bf Abstract}\\[2mm]
We present an overview of the options for diffraction implemented
in the general-purpose event generator \textsc{Pythia}~8 
\cite{Sjostrand:2014zea}. We
review the existing model for soft diffraction and present a new 
model for hard diffraction. Both models use the Pomeron approach 
pioneered by Ingelman and Schlein, factorising the diffractive cross 
section into a Pomeron flux and a Pomeron PDF, with several choices for 
both implemented in \textsc{Pythia}~8. The model of hard
diffraction is implemented as a part of the multiparton
interactions (MPI) framework, thus introducing a dynamical gap
survival probability that explicitly breaks factorisation.
\end{minipage}
\end{center}

\vspace{\fill}

\phantom{dummy}

\clearpage

\pagestyle{plain}
\setcounter{page}{1}

\section{Introduction}
\label{intro}
Despite the great success of QCD many soft phenomena
still remain as mysteries because of their nonperturbative
origin. Amongst these lies soft hadron collisions,
in which one or both of the hadrons survive the collision. In
this regime we cannot predict the collisions or their rate with
our usual perturbative framework of Feynman diagrams and
factorisation theorems, as these require a breakup of the hadrons
into their constituents, which then collide. We must therefore
base our work on phenomenological models, many of which has
sprung out of theories invented before QCD, such as S-matrix
theory and the related Regge theory. 

In Regge theory, poles in the plane of complex spin, $\alpha$,
can be seen as hadronic resonances. These appear to lie on linear
trajectories, $\alpha(t) = \alpha(0) + \alpha't$, of which the
most important for high-energy applications is the Pomeron
($\mathbb{P}$) trajectory, with its $\alpha(0)>1$ explaining the
rise of the total cross section. The Pomeron must be a colour
singlet carrying the quantum numbers of the vacuum, as it should
be able to describe elastic scattering of the hadrons. From a
more modern viewpoint it would mainly consist of gluons, but a
quark content could also be envisioned -- as long as this content
come in quark-antiquark pairs, in order to preserve the singlet 
nature of the $\mathbb{P}$. Having defined the state and its
possible partonic content, one can perform a topological
expansion similar to that of perturbative QCD, where the simplest
possible exchange is a single-$\mathbb{P}$ one, describing elastic
scattering. More complex processes can be constructed through the
triple-$\mathbb{P}$ vertex, resulting in the different diffractive
topologies. We generally work with three different diffractive
processes, all involving the triple-$\mathbb{P}$ vertex; the
single-diffractive (SD) processes (one surviving hadron and one
triple-$\mathbb{P}$ vertex), double-diffractive (DD) processes (no
surviving hadrons, two triple-$\mathbb{P}$ vertices) and
central-diffractive (CD) processes (two surviving hadrons and two
triple-$\mathbb{P}$ vertices). In this paper we focus on the SD
processes, being the simplest possible configuration and the one
with the largest cross section.

Ingelman and Schlein proposed a model \cite{Ingelman:1984ns} in which
the $\mathbb{P}$ can be viewed as a hadronic state with partonic
content. This opened up the possibility for $\mathbb{P}$ parton
distribution functions (PDFs) to be combined with the probability
of extracting a $\mathbb{P}$ from the colliding hadron, denoted
the $\mathbb{P}$ flux. Hence the diffractive system could be
envisioned as a hadron-hadron collision at reduced energy, which
opened up the possibility for modelling the diffractive processes
with existing hadron-hadron event generators.  

A model for diffractive systems should be able to
describe all aspects of these collisions, such as differential 
cross sections, one-particle distributions and global event 
characteristics. The framework used in \textsc{Pythia}~8
describes all of these aspects in detail. As rapidity gaps are 
crucial for diffraction, this framework does not permit MPIs 
in the $\mathrm{pp}$ system, as these would fill up the 
would-have-been rapidity gap. Thus we introduce the concept 
of rapidity gap survival probability (RGSP) (cite Bjorken), 
a concept unique to hadron-hadron collisions, given credibility 
by the lower observed rate for diffraction at the Tevatron as 
expected from HERA flux and PDF determinations.

\section{Soft diffraction in \textsc{Pythia}~8}
\label{sec-1}
The soft diffraction framework available in \textsc{Pythia}~8 was
originally developed for its predeccesor, \textsc{Pythia}~6 
\cite{Sjostrand:2006za}, but
rewritten and expanded for the new version \cite{Navin:2010kk}. 
The total hadronic
cross section is calculated using the Donnachie-Landshoff \cite{Donnachie:1992ny}
parametrisation, including both a Pomeron and Reggeon term. The
elastic and diffractive cross sections are calculated using the
Schuler-Sj\"ostrand model \cite{Schuler:1993wr}, 
with the nondiffractive cross
section inferred from the above. 

The Schuler-Sj\"ostrand model is based on Regge theory, giving an
exponential $t$-dependence and an approximate $dM^2/M^2$ mass
dependence, with fudge factors introduced in order to dampen the
cross sections close to the kinematical limits, as well as dampen
the DD cross section in regions where the two diffractive systems
overlap. Other $\mathbb{P}$-flux models have also been
implemented, all of which have been expanded to include both DD
and CD where needed \cite{PythiaWeb}. The subsequent evolution and
hadronisation of the $\mathbb{P}\mathrm{p}$ system is separated
from the $\mathbb{P}$-parametrisation, hence being the same for
almost all models.

\subsection{Low-mass soft diffraction}

In the low-mass regime, $M\leq10$ GeV, the energies are not
sufficiently high to apply a perturbative framework to the
$\mathbb{P}\mathrm{p}$ subsystem. The framework is setup as an
interaction, where a $\mathbb{P}$ has ``kicked out'' a quark or a
gluon from the incoming hadron. In the former case a colour
string is stretched between the ``free'' quark and the diquark
remnant inside the proton. In the latter case a colour string is
stretched between one valence quark inside the hadron through 
the ``free'' gluon and back to the diquark in the remnant,
resulting in a hairpin-like structure. The probability for the
$\mathbb{P}$ to interact in either of the two ways is
mass-dependent and tunable, $\mathcal{P}(q)/\mathcal{P}(g) =
N/M^p$, set up such that the gluons dominate at higher masses.
Since this is a nonperturbative framework, no additional
evolution of the colour strings is applied. Instead the system
is hadronised directly using the Lund string fragmentation model
\cite{Andersson:1983ia}, 
thus giving rise only to low-$p_{\perp}$ activity in the
diffractive system.

\subsection{High-mass soft diffraction}

In the high-mass regime, $M>10$ GeV, energies are sufficiently
high to use a perturbative framework. The high-mass regime is
smoothly connected to the low-mass one, with the fraction of
perturbative events gradually increasing from the 10 GeV starting
point and dominating at diffractive masses above 20 GeV. In this
regime the $\mathbb{P}$ is viewed as having a partonic content. 
Thus once $M$ and $t$ have
been selected according to the model of the $\mathbb{P}$ flux,
the $\mathbb{P}\mathrm{p}$ subsystem is set up and a semi-hard
perturbative $2\rightarrow2$ QCD process is selected by the MPI
framework as the hardest collision. The subsystem is then further
evolved using the interleaved parton evolution of
\textsc{Pythia}~8, including both MPI as well as initial and
final state showers (ISR, FSR), using the $\mathbb{P}$ PDFs on
the diffractive side. 

The MPI acitivity in the $\mathbb{P}\mathrm{p}$ system has been
tuned to give approximately the same amount of activity as
nondiffractive events of the same mass by introducing an
effective total $\mathbb{P}\mathrm{p}$ cross section of 10 mb.
This is a tunable parameter, and can be made to depend on the
mass of the diffractive system, just as the total $\mathrm{pp}$
cross section does. The evolution results in a larger number of colour
strings in the event, all hadronised with the Lund string
fragmentation model, giving rise to low- to
high-$p_{\perp}$ activity in the event. 

Although largely successful, the model for soft diffraction in
\textsc{Pythia}~8 has minor issues. Not all aspects of data are
described by the default model and settings, including both
differential cross sections and particle spectra. A full retune
could improve some of these discrepancies, but a more flexible 
shape for the $\mathbb{P}$ flux may be needed
in order to describe data better. Such studies are planned 
for the near future. 

\section{Hard diffraction in \textsc{Pythia}~8}

The model described in the previous section allows for
$2\rightarrow2$ QCD processes at all $p_{\perp}$ scales, but is
primarily intended for lower values of $p_{\perp}$. For truly
hard diffractive processes, the new model for hard diffraction
\cite{Rasmussen:2015qgr} was developed, not only for 
high-$p_{\perp}$ jets, but also allowing for $W^{\pm}, Z^0, H$ etc. 

The model is based on the assumption that the proton PDF can be
split into a diffractive and a nondiffractive part, 
\begin{align}
f_{i/\mathrm{p}}(x,Q^2) = f_{i/\mathrm{p}}^{\mathrm{ND}}(x,Q^2) +
f_{i/\mathrm{p}}^{\mathrm{D}}(x,Q^2),
\end{align}
with the diffractive part being described using the factorisation
approach of Ingelman and Schlein,
\begin{align}
f_{i/\mathrm{p}}^{\mathrm{D}}(x, Q^2) &=  \int_0^1 \mathrm{d} x_{\mathbb{P}} \, 
  f_{\mathbb{P}/\mathrm{p}}(x_{\mathbb{P}})\,  \int_0^1 \mathrm{d} x' \, 
  f_{i/\mathbb{P}}(x', Q^2) \, \delta(x - x_{\mathbb{P}} x') \nonumber \\
&=  \int_x^1  \frac{\mathrm{d} x_{\mathbb{P}}}{x_{\mathbb{P}}} \,  
  f_{\mathbb{P}/\mathrm{p}}(x_{\mathbb{P}}) \, f_{i/\mathbb{P}} \left( 
  \frac{x}{x_{\mathbb{P}}}, Q^2 \right) ~,
\end{align}
where $f_{\mathbb{P}/\mathrm{p}}(x_{\mathbb{P}}) =
\int_{t_{\mathrm{min}}}^{t_{\mathrm{max}}}\mathrm{d} t \, 
f_{\mathbb{P}/\mathrm{p}}(x_{\mathbb{P}}, t)$, as $t$ for the
most part is not needed.

The probability for side $A,B$, to be diffractive is then
given by the ratio of diffractive to inclusive PDFs,
\begin{align}
\label{Eq:Prob}
\mathcal{P}_A^{\mathrm{D}} &= \frac{f_{i/B}^{\mathrm{D}}(x_B, Q^2)}
{f_{i/B}(x_B, Q^2)}~~~\mathrm{for}~~~AB \to XB~, \nonumber \\
\mathcal{P}_B^{\mathrm{D}} &= \frac{f_{i/A}^{\mathrm{D}}(x_A, Q^2)}
{f_{i/A}(x_A, Q^2)}~~~\mathrm{for}~~~AB \to AX~.
\end{align}

\begin{figure}
\centering
\includegraphics[scale=0.4]{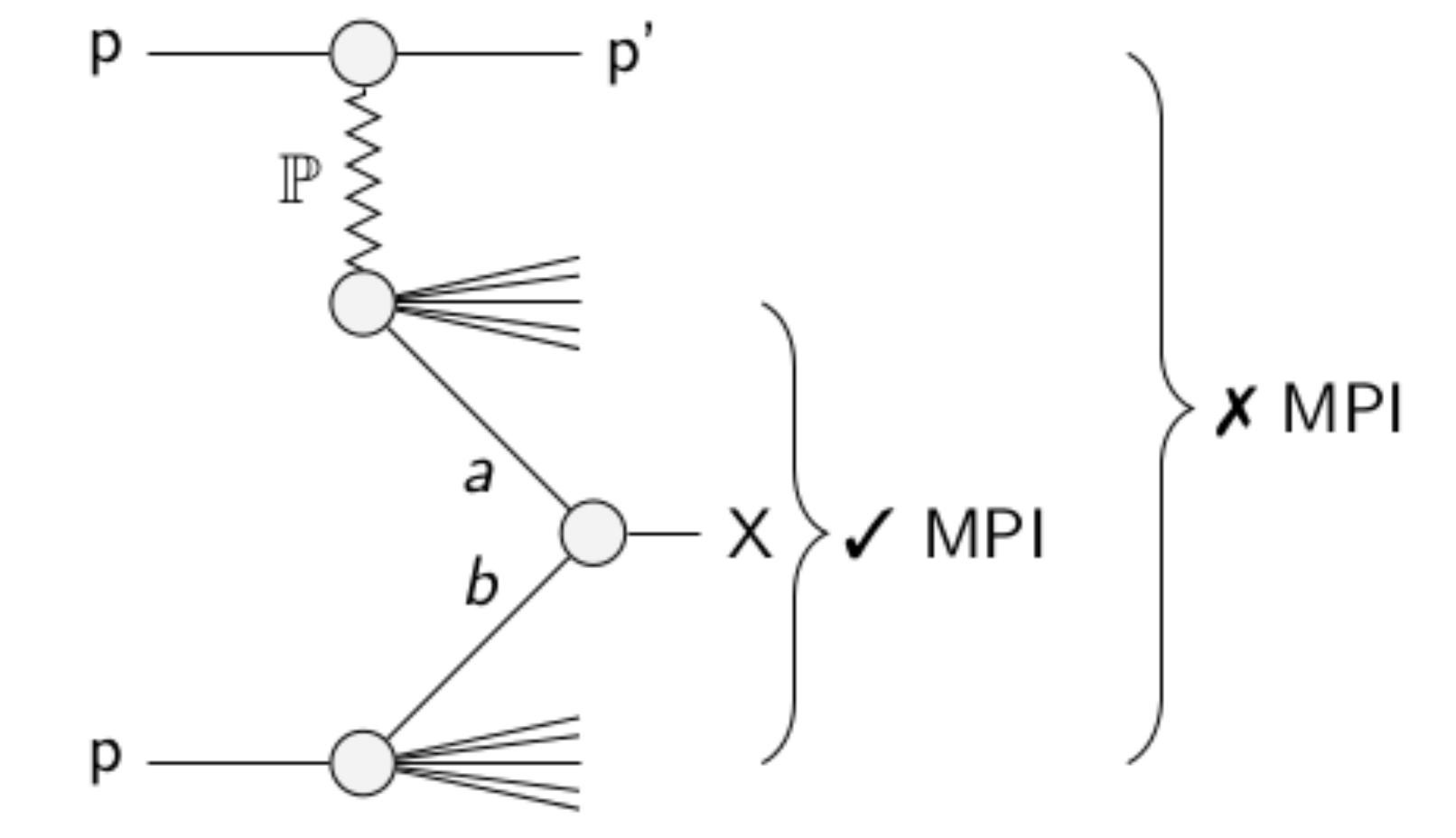}
\caption{\label{Fig:DynGapSur} The dynamical gap survival implemented 
in \textsc{Pythia}~8. The model do not allow for MPIs in the
$\mathrm{pp}$ system, but allows for additional MPIs in the
$\mathbb{P}\mathrm{p}$ system.}
\end{figure}

The model also implements a dynamical gap survival, cf.\
Fig.~\ref{Fig:DynGapSur}. On an
event-by-event basis we evaluate the possibility for additional
MPIs in the $\mathrm{pp}$ system, and if no further MPIs are
found, then the event is diffractive. Thus initially
eq.~(\ref{Eq:Prob}) has no consequence, with all events still
handled as being nondiffractive. Only if no additional MPIs are 
found is the event classified as diffractive, and the
$\mathbb{P}\mathrm{p}$ system set up. A full evolution is then
performed in this subsystem, along with the hadronisation of the
colour strings in the event. 
At this point, the nondiffractive events can be discarded
if an exclusively diffractive sample is wanted, otherwise they
can be kept for an inclusive sample consisting of both non- and
single-diffractive events.

\begin{table}
\begin{center}
\caption{\label{Tab:Fractions}
Diffractive fractions for the $2\rightarrow2$ QCD processes 
with $p_{\perp} > 20$ GeV in $\mathrm{p}\mathrm{p}$ collisions 
at $\sqrt{s}=8$ TeV obtained with \textsc{Pythia}~8.
The samples have been produced without any phase-space cuts.
$\mathbb{P}$ parametrizations not cited in the text are: H1 Fit A
NLO PDF \cite{Aktas:2006hy} and H1 Fit A flux 
\cite{Aktas:2006hy, Aktas:2006hx}}
\begin{tabular}{lll}
\hline
\begin{tabular}{c}$\mathbb{P}$ PDF \\ $\mathbb{P}$ flux \end{tabular}
& PDF selection & MPI selection\\
\hline
H1 Fit B LO & & \\
Schuler-Sj\"ostrand & (14.33 $\pm$ 0.11) \% & (0.98 $\pm$ 0.03) \%\\ 
H1 Fit B LO & & \\
MBR                 & (14.79 $\pm$ 0.11) \% & (0.96 $\pm$ 0.03) \%\\ 
H1 Jets & & \\
Schuler-Sj\"ostrand & (13.70 $\pm$ 0.11) \% & (0.92 $\pm$ 0.03) \%\\ 
H1 Fit A NLO & & \\
H1 Fit A            & (20.55 $\pm$ 0.13) \% & (1.35 $\pm$ 0.04) \%\\ 
H1 Fit B LO & & \\
H1 Fit A            & (18.49 $\pm$ 0.12) \% & (1.32 $\pm$ 0.04) \%\\ 
\hline
\end{tabular}
\end{center}
\end{table}

The dynamical gap survival introduces an additions suppression of
the diffractive events, such that the total probability for
hard diffraction drops from $\sim10$\% to $\sim1$\%. In
Table~\ref{Tab:Fractions} we show the number of events passing
either the probabilistic criterion of eq.~(\ref{Eq:Prob})
(PDF selection) or both the probabilistic and the dynamical gap
survival (MPI selection) for a variety of
the $\mathbb{P}$ fluxes and PDFs available in \textsc{Pythia}~8.
These fractions are very model dependent, of course. They depend
both on the $\mathbb{P}$ parametrization and on the free
parameters of the MPI framework, although  many of the
distributions tend to be mainly driven by only one of them. 

\begin{figure}
\begin{minipage}[t]{0.5\textwidth}
\centering
\includegraphics[scale=0.3]{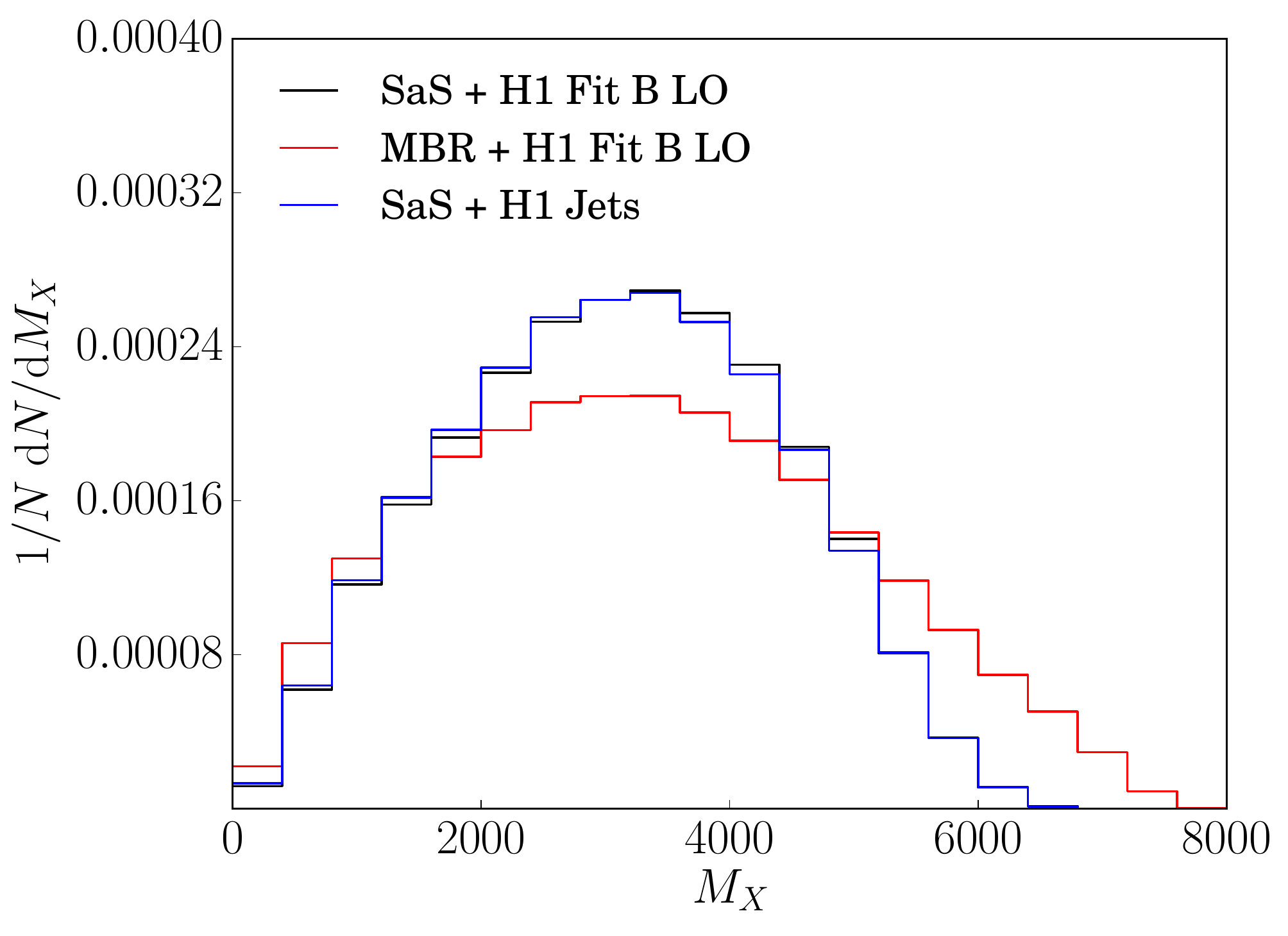}\\
(a)
\end{minipage}
\hfill
\begin{minipage}[t]{0.5\textwidth}
\centering
\includegraphics[scale=0.3]{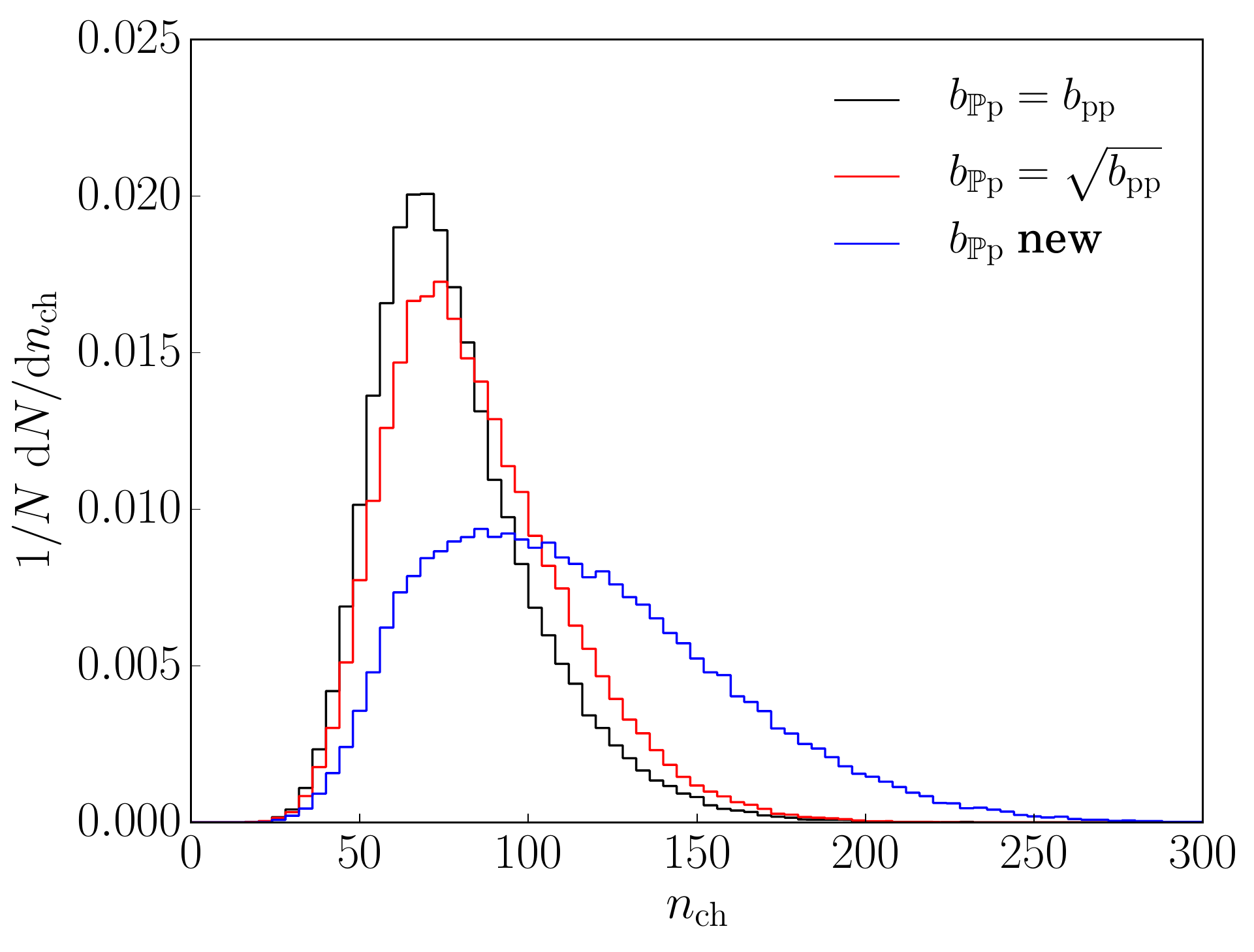}\\
(b)
\end{minipage}\\[2mm]
\caption{\label{Fig:PomDistVal}
(a) The mass of the diffractive system, $M_X$, when changing
$\mathbb{P}$ parametrization. 
(b) The number of charged particles in the $\mathbb{P}\mathrm{p}$
system, when changing the size of the $\mathbb{P}$.}
\end{figure}

Distributions affected by the modelling of the $\mathbb{P}$
flux and PDF are e.g.\ the chosen $x_{\mathbb{P}}$ value,
hence the mass of the diffractive system, as well as the value 
of the squared momentum transfer, $t$ and through that the
angle at which the proton is scattered. The mass of the
diffractive system is shown in Fig.~\ref{Fig:PomDistVal}a, 
using the following parametrizations: The Schuler-Sj\"ostrand 
flux (SaS) \cite{Schuler:1993wr}, the Minimum Bias Rockefeller 
lux (MBR) \cite{Ciesielski:2012mc}, the H1 Fit B LO PDF
\cite{Aktas:2006hy} and the H1 Jets PDF \cite{Aktas:2007bv}. It
is observed that the variation of the $\mathbb{P}$ flux affects
the distribution more than the variation of the $\mathbb{P}$ PDF.

Those affected by the MPI parameters are several particle 
distributions such as multiplicities and $p_{\perp}$ spectra. One
of these is shown in Fig.~\ref{Fig:PomDistVal}b, where the
transverse matter profile of the $\mathbb{P}$ is varied. In the
end, what really matters is the convolution of the matter profile
of the $\mathbb{P}$ and the $\mathrm{p}$, resulting in an overlap
function, from where the impact parameter $b$, describing the 
amount of overlap between the $\mathbb{P}$ and $\mathrm{p}$, is
chosen. We have implemented
three options: 
\begin{itemize}
\item $b_{\mathbb{P}\mathrm{p}}=b_{\mathrm{pp}}$,
implying that the $\mathbb{P}$ and $\mathrm{p}$ are equally big
and have equal matter profiles. 
\item $b_{\mathbb{P}\mathrm{p}}=\sqrt{b_{\mathrm{pp}}}$,
motivated if the $\mathbb{P}$ is a point particle. The variation
of the overlap function with $b$ is then that of one proton, not
two, giving the square root of the normal variation. 
\item The final option implemented is picking a completely new
$b_{\mathbb{P}\mathrm{p}}$, ie.\ not specifying the size of the
$\mathbb{P}$ or its matter profile. 
\end{itemize}
The variation of the size gives different
distributions, the latter definition giving much larger
multiplicities than the two first (and probably more realistic)
cases. 

Also the gap survival probability is highly sensitive to the
parameters of the MPI model, highlighted in
Table~\ref{Tab:pT0ref}, where the regulator of the
$2\rightarrow2$ QCD cross section, $p_{\perp 0}^{\mathrm{ref}}$
has been varied with $\pm0.5$ GeV, resulting in a factor of two
in the diffractive fractions. This major $p_{\perp 0}^{\mathrm{ref}}$ 
dependence holds also for many nondiffractive event properties, 
however; keeping everything else fixed even a variation of 
$\pm 0.1$~GeV would be unacceptable. It would be possible to perform a
dedicated diffractive tune, but the same parameters would not be
able to describe nondiffractive data. Such
a retune is considered.

\begin{table}
\begin{center}
\caption{\label{Tab:pT0ref}
Diffractive fractions for the $2\rightarrow2$ QCD processes with 
$p_{\perp}>20$ GeV in $\sqrt{s}=8$ TeV $\mathrm{pp}$ collisions. 
\textsc{Pythia} is run with the Schuler-Sj\"ostrand flux and 
the H1 Fit B LO PDF.}
\begin{tabular}{lll}
\hline
& PDF selection & MPI selection\\
\hline
$p_{\perp 0}^{\mathrm{ref}}=1.78$ & (14.50 $\pm$ 0.11) \% & (0.39 $\pm$ 0.02) \% \\
$p_{\perp 0}^{\mathrm{ref}}=2.28$ & (14.33 $\pm$ 0.11) \% & (0.98 $\pm$ 0.03) \%\\ 
$p_{\perp 0}^{\mathrm{ref}}=2.78$ & (14.19 $\pm$ 0.11) \% & (2.00 $\pm$ 0.04) \% \\
\hline
\end{tabular}
\end{center}
\end{table}

\section{Conclusions}

We have presented the two frameworks for diffraction present in
the general-purpose event generator \textsc{Pythia}~8. The soft
diffraction framework allows for QCD processes at various
$p_{\perp}$ scales, although primarily intended for
low-$p_{\perp}$ interactions. It gives a decent description of
soft diffraction, although comparisons to data show room for
improvement. The new model for hard diffraction allows for
non-QCD and very high-$p_{\perp}$ QCD processes in diffractive
systems. The effects of the $\mathbb{P}$ parametrizations as well
as the free parameters of the MPI framework have been explored.
Future work includes a retuning of the free parameters in both
models and developing a new parametrization of the $\mathbb{P}$
flux, in order to obtain a better description of data.

\section*{Acknowledgements}

Work supported by the MCnetITN FP7 Marie Curie Initial 
Training Network, contract PITN-GA-2012-315877.

\end{document}